\documentstyle[12pt,epsf,axodraw]{article}
\makeatletter

\@addtoreset{equation}{section}
\makeatother
\begin{document}
\baselineskip 0.8cm

\newcommand{\gsim}{ \mathop{}_{\textstyle \sim}^{\textstyle >} }
\newcommand{\lsim}{ \mathop{}_{\textstyle \sim}^{\textstyle <} }
\newcommand{\vev}[1]{ \left\langle {#1} \right\rangle }
\newcommand{\EV}{ {\rm eV} }
\newcommand{\KEV}{ {\rm keV} }
\newcommand{\MEV}{ {\rm MeV} }
\newcommand{\GEV}{ {\rm GeV} }
\newcommand{\TEV}{ {\rm TeV} }
\newcommand{\DS}{\displaystyle}
\def\tr{\mathop{\rm tr}\nolimits}
\def\Tr{\mathop{\rm Tr}\nolimits}
\def\Re{\mathop{\rm Re}\nolimits}
\def\Im{\mathop{\rm Im}\nolimits}
\def\simgt{\mathop{>}\limits_{\displaystyle{\sim}}}
\def\simlt{\mathop{<}\limits_{\displaystyle{\sim}}}
\setcounter{footnote}{0}

\begin{titlepage}

\begin{flushright}
OCHA-PP-151\\
AJC-HEP-33\\
\end{flushright}

\vskip 2cm
\begin{center}
{\large \bf  Production of CP-even and CP-odd Higgs bosons \\
at Muon colliders}
\vskip 1.2cm
Eri Asakawa$^*$, Akio Sugamoto$^*$, and Isamu Watanabe$^{**}$

\vskip 0.4cm

$^{*}$ {\it Department of Physics and \\
Graduate School of Humanities and Sciences, Ochanomizu University \\
         Tokyo 112-8610, Japan}\\
$^{**}$ {\it Akita Keizaihoka University \\
         Akita 010-8515, Japan}
\vskip 1.5cm

\abstract{
In the $s$-channel Higgs-boson-exchange processes,
the interference between the amplitudes for
CP-even and CP-odd Higgs bosons is sizable,
if the helicities of initial and final particles are 
properly fixed and if the mass difference between 
these bosons is not much larger than their decay widths.
We discuss this interference effect 
in the process $\mu^+ \mu^-$ $\rightarrow t\overline{t}$.
Examining the effects gives us information
on CP-parity for Higgs bosons and 
on the sign of a product of the coupling constants
for $H\mu^-\mu^+$, $Ht\overline{t}$,
$A\mu^-\mu^+$ and $At\overline{t}$ vertices.
The feasibility of observing the interference effect
in future muon colliders
is evaluated in the framework of the minimal supersymmetric
extension of the standard model as an example.
}

\end{center}
\end{titlepage}

\renewcommand{\thefootnote}{\arabic{footnote}}
\setcounter{footnote}{0}

%
%
%
%
\section{Introduction}

The standard model (SM) of particle physics predicts one physical
neutral CP-even Higgs boson.
It can be extended by increasing Higgs fields.
In models with more than one Higgs doublets
there are extra two neutral and two charged physical Higgs bosons
for each additional doublet.
If CP is a good symmetry, one neutral boson is CP-even ($H$) and
the other is CP-odd ($A$).
Searches for the Higgs bosons and 
precise measurements of their properties are indispensable for 
understanding the mechanism of electroweak symmetry breaking
and knowing which model is realized in nature.

A muon collider is one of the ideal machines to look for 
the Higgs bosons~\cite{BBGH-PR, higgs-s-channel},
where $\mu^+ \mu^-$ pairs are directly annihilated into 
$H$ and $A$ bosons. The muon beams can be polarized~\cite{MPOL}.  
The feasibility of detecting the Higgs bosons and measuring their properties, 
such as masses, total widths and 
decay branching fractions has been studied~\cite{BBGH-PR,
higgs-s-channel, precise-measure, loopCP, GGP, AS}.
They have covered the special cases where $H$ and $A$ bosons are 
distributed overlapping their resonances
~\cite{BBGH-PR} which are often expected 
in the minimal supersymmetric extension of the SM (MSSM)
and where mixing between $H$ and $A$ bosons occurs due to
loop-induced CP violation~\cite{loopCP}.
Moreover, determining the CP-nature of Higgs bosons has been
also discussed
by using initial muon polarizations~\cite{GGP}.

We here take into consideration of overlapped resonances,
especially concentrating on interference between the resonances.
As was pointed out recently in Ref.~\cite{AKSW}
where the process $\gamma\gamma \rightarrow t \overline{t}$
is studied, the amplitudes of $s$-channel exchanges for
$H$ and $A$ bosons can sizably interfere 
with each other,
if the helicities of the initial and final particles are
fixed properly and if the mass difference between
these Higgs bosons is at most of the same order
as their decay widths.
This interference effect disappears 
if the helicities of both initial and final particles
are not fixed.
Similar effects are also expected in the process 
$\mu^+ \mu^- \rightarrow t \overline{t}$.

In this paper, 
we discuss the interference effects on the cross section
of the process $\mu^+ \mu^-$ $\rightarrow t\overline{t}$
assuming definite helicities of the initial and final particles.
Although the process $\mu^+ \mu^- \rightarrow t \overline{t}$
is also generated by
$s$-channel $\gamma$- and $Z$-exchange diagrams,
their contributions
are small under the helicity combinations which induce 
the $s$-channel $H$- and $A$-exchange amplitudes.
The interference effects are measured
by the difference between the cross sections
for various helicity selections.
We estimate an asymmetry between these cross sections.
Examining the effects gives us information on CP-parity
for the Higgs bosons. 
In addition, the asymmetry
provides information about the sign of a
product of the coupling constants for $H\mu^-\mu^+$, $Ht\overline{t}$,
$A\mu^-\mu^+$ and $At\overline{t}$ vertices.
Since this sign depends on a model, 
we can judge a type of the model from a viewpoint of the coupling constants.
An example is given in the MSSM which contains two Higgs doublets.

This paper is organized as follows.
In Sect.~2 we obtain helicity amplitudes of the process 
$\mu^+\mu^- \rightarrow t \overline{t}$. 
In Sect.~3 interference effects are discussed and an asymmetry
between the cross sections is defined. 
In Sect.~4 numerical estimates of the cross sections
and the asymmetry are given.
The degree of polarization of muon beams and 
a method of helicity observation of final top pairs are
also considered. We give conclusions in the last section.

\section{Helicity Amplitudes}

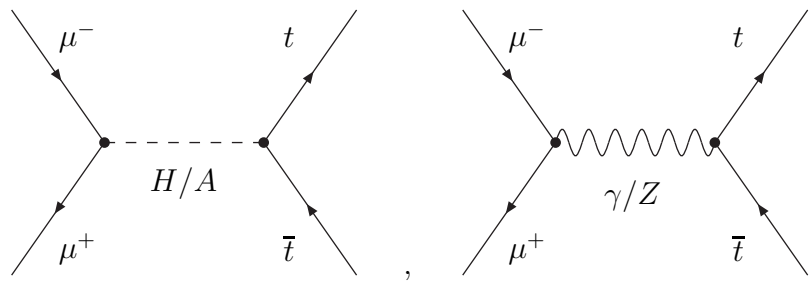
\begin{figure}[h]
\section*{Figures}
\begin{center}\begin{picture}(300,100)(0,0)
\SetColor{Black}
\ArrowLine(0,100)(35,50) \Text(25,90)[]{$\mu^-$} \Vertex(35,50){2}
\ArrowLine(35,50)(0,0)   \Text(25,10)[]{$\mu^+$}
\DashLine(35,50)(95,50){4} \Text(65,35)[]{$H$/$A$}
\ArrowLine(95,50)(130,100) \Text(105,90)[]{$t$}
\ArrowLine(130,0)(95,50) \Text(105,10)[]{$\overline{t}$} 
\Vertex(95,50){2}
\Text(150,0)[]{,}
\ArrowLine(170,100)(205,50) \Text(195,90)[]{$\mu^-$} 
\Vertex(205,50){2}
\ArrowLine(205,50)(170,0) \Text(195,10)[]{$\mu^+$}
\Photon(205,50)(265,50){5}{6} \Text(235,30)[]{$\gamma$/$Z$}
\ArrowLine(265,50)(300,100) \Text(275,90)[]{$t$}
\ArrowLine(300,0)(265,50) \Text(275,10)[]{$\overline{t}$}
\Vertex(265,50){2}
\end{picture}
\caption{The diagrams of the process $\mu^+\mu^-
\rightarrow t\overline{t}$ around the mass poles of
$H$ and $A$ bosons.} \label{dia}
\end{center}
\end{figure}

The process $\mu^+ \mu^- \rightarrow t \overline{t}$
receives contributions from the diagrams in which
$H$, $A$, $\gamma$ and $Z$ are exchanged as shown in Fig.~\ref{dia}.
We express the helicity amplitudes for these diagrams as
\begin{eqnarray}
{\cal M}^{\Lambda \overline{\Lambda} \lambda \overline{\lambda}}_{H},
{\cal M}^{\Lambda \overline{\Lambda} \lambda \overline{\lambda}}_{A},
{\cal M}^{\Lambda \overline{\Lambda} \lambda \overline{\lambda}}_{\gamma},
{\cal M}^{\Lambda \overline{\Lambda} \lambda \overline{\lambda}}_{Z}.
\end{eqnarray}
The superscripts $\Lambda$ and $\overline{\Lambda}$ denote
the initial $\mu^-$ and $\mu^+$ helicities, while 
$\lambda$ and $\overline{\lambda}$
the final $t$ and $\overline{t}$ helicities in the center-of-mass frame.
Defining $\lambda_i$, $\lambda_f$ as $\Lambda-\overline{\Lambda}$,
$\lambda-\overline{\lambda}$, respectively,
the amplitudes are given by
\begin{eqnarray}
{\cal M}_H^{\Lambda \overline{\Lambda} \lambda \overline{\lambda}}
&= & -4\Lambda\lambda a_{\mu} a_t g^2 \frac{m_{\mu} m_t}{m_W^2}
\beta_{\mu}\beta_t \frac{\DS s}{\DS s-m_H^2+im_H\Gamma_H}
\delta_{\lambda_i0} \delta_{\lambda_f0} , \label{H}\\
{\cal M}_A^{\Lambda \overline{\Lambda} \lambda \overline{\lambda}}
&= & -b_\mu b_t g^2 \frac{m_{\mu} m_t}{m_W^2} 
\frac{\DS s}{\DS s-m_A^2+im_A\Gamma_A}
\delta_{\lambda_i0} \delta_{\lambda_f0}, 
\label{A} \\
{\cal M}_{\gamma}^{\Lambda \overline{\Lambda} \lambda \overline{\lambda}}
 &=& \overline{\Lambda}\,\overline{\lambda}\frac{16}{3}\pi\alpha_{QED}
K_{\mu} K_t d_{\lambda_i \lambda_f}^1 
\label{g}\\
{\cal M}_Z^{\Lambda \overline{\Lambda} \lambda \overline{\lambda}}
 &=& 
\frac{\pi\alpha_{QED}}{2\sin^2\theta_W\cos^2\theta_W} K_{\mu} K_t 
\left[ F_0 \delta_{\lambda_i 0} \delta_{\lambda_f 0} 
 - \overline{\Lambda}\,\overline{\lambda} F_{\mu} F_t 
   d_{\lambda_i \lambda_f}^1 \right] \frac{\DS s}{\DS s - m_Z^2} , 
\label{Z} \\
K_{\mu} & = & \delta_{\lambda_i 0} \sqrt{\frac{2}{s}}m_{\mu}+|\lambda_i| , \quad 
K_t     \ = \ \delta_{\lambda_f 0} \sqrt{\frac{2}{s}}m_t+|\lambda_f| , \\
F_0     & = & \frac{\DS s - m_Z^2}{m_Z^2} , \nonumber \\
F_{\mu} & = & -1 + 4 \sin^2\theta_W + \lambda_i \beta_{\mu} , \quad 
F_t     \ = \ +1 -\frac{8}{3}\sin^2\theta_W - \lambda_f \beta_t , 
\end{eqnarray}
where $g$ is the weak coupling constant; 
$m_{\mu}$, $m_t$, $m_W$ and $m_Z$ are the masses of muon, top quark, 
$W$ boson and $Z$ boson; 
$\beta_t$ and $\beta_{\mu}$ are the velocities of the top quarks
and the muons in the center-of-mass frame;
$s$ is the collision energy-squared;
and $d$ is the Wigner's $d$ function.  
The masses and the total decay widths 
of Higgs bosons are denoted by $m_{H,A}$ and $\Gamma_{H,A}$.  
The coefficients 
$a_\mu$, $a_t$, $b_\mu$ and $b_t$ are the coupling constants 
for $H\mu^-\mu^+$, $Ht\overline{t}$, $A\mu^-\mu^+$
and $At\overline{t}$ vertices, respectively. 
In the framework of the MSSM, they are expressed as
\begin{eqnarray}
a_\mu = -\frac{1}{2}\frac{\cos\alpha}{\cos\beta},~
~a_t = -\frac{1}{2} \frac{\sin\alpha}{\sin\beta},~
~b_\mu = \frac{1}{2} \tan\beta,~
~b_t = \frac{1}{2} \cot\beta. \label{MSSMcup}
\end{eqnarray}
Here $\alpha$ is the mixing angle of the two CP-even Higgs bosons
and $\tan\beta$ is the ratio of the vacuum expectation values of
two Higgs doublets.
Here $-4 a_{\mu} a_t$ is almost unity within the 
accuracy of 13{\%} for $\tan\beta$ = 1---100 in the MSSM, which is 
comparable with $4 b_{\mu} b_t$ = 1.  

The Higgs-exchange diagrams
can contribute only for $\lambda_i=\lambda_f=0$. 
The absolute values of ${\cal M}_{\gamma}$ and ${\cal M}_Z$
in this case are proportional to $m_{\mu}$, and 
negligibly small around the mass poles 
of the Higgs bosons if the masses of the Higgs bosons are 
far above $m_Z$, as will be seen in the subsection 3.3.

\section{Interference Effects}

The signs of Higgs-exchange amplitudes
change under CP transformation.
Expressing ${\cal M}_{H, A}^{RRRR}$ as ${\cal M}_{H, A}$,
where $R$ ($L$) means the helicity $+1/2$ ($-1/2$),
the helicity dependence of the Higgs-exchange amplitudes is summarized
in Table~\ref{helamp}.
With $|RR\rangle$ and $|LL\rangle$ being
spin-zero states of $t_R \overline{t}_R$
or $\mu^+_R \mu^-_R$, and $t_L \overline{t}_L$
or $\mu^+_L \mu^-_L$ system, respectively, 
these states are interchanged under CP transformation,
\begin{eqnarray}
{\cal CP} |RR\rangle & = & - |LL\rangle \ , \nonumber \\
{\cal CP} |LL\rangle & = & - |RR\rangle \ .
\label{parity}
\end{eqnarray}
For example, ${\cal M}_H^{RRRR}$ and ${\cal M}_H^{RRLL}$
are related by CP transformation of the final 
$t \overline{t}$ system and have the same absolute value of 
the amplitudes with the different signs. 
As for ${\cal M}_A$, the odd contribution of CP transformation
for the final $t \overline{t}$ state is cancelled by
the transformation of the CP-odd Higgs boson.

With the helicities fixed, the interference term of the amplitudes
for the Higgs bosons, 
$\Delta\sigma^{\Lambda\overline{\Lambda}\lambda\overline{\lambda}}$, 
is given by
\begin{eqnarray}
\Delta\sigma^{\Lambda\overline{\Lambda}\lambda\overline{\lambda}} 
&=& \frac{N_c}{8\pi s} \frac{\beta_t}{\beta_{\mu}}
{\cal R}e \bigl[ 
{\cal M}_{\phi_1}^{\Lambda\overline{\Lambda}\lambda\overline{\lambda}} \cdot 
{\cal M}_{\phi_2}^{* \Lambda\overline{\Lambda}\lambda\overline{\lambda}} \bigr]
\end{eqnarray}
where $N_c$ is the color factor of the top quark,
$\phi_1$ and $\phi_2$ denote the relevant two Higgs bosons.
The interference between the amplitudes for two Higgs bosons is
manifestly dependent on the relative CP-parity of the Higgs bosons.
According to Tabel~\ref{helamp} , the terms 
$\Delta\sigma^{\Lambda\overline{\Lambda}\lambda\overline{\lambda}}$
for different helicity states are related to each other.
For $\phi_1$ = $H$ and $\phi_2$ = $A$, 
\begin{eqnarray}
\Delta\sigma^{RRRR} = \Delta\sigma^{LLLL} = 
-\Delta\sigma^{RRLL} = -\Delta\sigma^{LLRR}, 
\end{eqnarray}
while for $\phi_1$ = $\phi_2$ = $H$ or $\phi_1$ = $\phi_2$ = $A$
\begin{eqnarray}
\Delta\sigma^{RRRR} = \Delta\sigma^{LLLL} = 
\Delta\sigma^{RRLL} = \Delta\sigma^{LLRR}.
\end{eqnarray}
Therefore, the comparison of the interference terms 
between the different helicity states could provide a useful information
on the Higgs CP-parity.

As an index for the interference term between two Higgs bosons,
we define an asymmetry of the cross sections as
\begin{eqnarray}
{\cal A} & \equiv & \frac{\sigma^{RRRR}+\sigma^{LLLL}
-\sigma^{RRLL}-\sigma^{LLRR}}
{\sigma^{RRRR}+\sigma^{LLLL}+\sigma^{RRLL}+\sigma^{LLRR}}.\label{asym}
\end{eqnarray}
where
$\sigma^{\Lambda \overline{\Lambda} \lambda \overline{\lambda}}$
denotes
the cross section of $\mu^+ \mu^- \rightarrow t \overline{t}$
with fixed helicities which is given by
\begin{eqnarray}
\sigma^{\Lambda \overline{\Lambda} \lambda \overline{\lambda}}=
\frac{N_c}{32\pi s} \frac{\beta_t}{\beta_{\mu}}
\int_{-1}^{+1} d\cos\theta \ 
\biggl| {\cal M}^{\Lambda \overline{\Lambda} \lambda \overline{\lambda}}
\biggr| ^2,
\end{eqnarray}
with ${\cal M}^{\Lambda \overline{\Lambda} \lambda \overline{\lambda}}$
denoting the sum of relevant amplitudes.
$\theta$ is the scattering angle
of the top quarks in center-of-mass frame. 
If the cross section receives contributions dominantly from
the diagrams mediated by the two Higgs bosons, 
the cross section can be written as
\begin{eqnarray}
\sigma^{\Lambda \overline{\Lambda} \lambda \overline{\lambda}}=
\frac{N_c}{16\pi s} \frac{\beta_t}{\beta_{\mu}} \ \biggl\{
\bigl| {\cal M}_{\phi_1} \bigr| ^2
+ \bigl| {\cal M}_{\phi_2} \bigr| ^2 \biggr\} 
+ \Delta \sigma^{\Lambda \overline{\Lambda} 
                 \lambda \overline{\lambda}},
\end{eqnarray}
and the asymmetry becomes
\begin{eqnarray}
{\cal A} &=& 
\frac{ 2 {\cal R}e \bigl[ {\cal M}_H \cdot {\cal M}_A^* \bigr] } 
         { |{\cal M}_H|^2 + |{\cal M}_A|^2 }
~~~~~( \phi_1= H~ \mbox{and} ~\phi_2= A), 
\nonumber \\
&=& 0 ~~~~~~~~~~~~~~~~~~~~~~~~~
 ( \phi_1=\phi_2 = H~ \mbox{or} ~\phi_1=\phi_2= A).
\end{eqnarray}
Therefore,
observation of a non-vanishing value for ${\cal A}$ indicates 
that the two Higgs bosons have different CP-parities.

In addition, from the sign of ${\cal A}$,
we can also learn the sign of the product of the coupling constants
for $H\mu^-\mu^+$, $Ht\overline{t}$,
$A\mu^-\mu^+$ and $At\overline{t}$ vertices. From 
eq.~(\ref{H}) and eq.~(\ref{A}) ,
\begin{eqnarray}
{\cal R}e \bigl[ {\cal M}_H \cdot {\cal M}_A^* \bigr]
= (a_{\mu} \cdot a_t \cdot b_{\mu} \cdot b_t)
~\beta_{\mu} \beta_t ~g^4~\frac{m_{\mu}^2 m_t^2}{m_W^4} ~s^2 ~{\cal D}
\end{eqnarray}
where
\begin{eqnarray}
{\cal D} \equiv
\frac{(s-m_H^2)(s-m_A^2)+m_H m_A \Gamma_H \Gamma_A}
{[(s-m_H^2)^2+m_H^2\Gamma_H^2][(s-m_A^2)^2+m_A^2\Gamma_A^2]}.
\end{eqnarray}
The sign of ${\cal A}$ is coincident with
the sign of $(a_{\mu} \cdot a_t \cdot b_{\mu} \cdot b_t)$
in the region of ${\cal D} \ge 0$ 
and is opposite to 
in the region of ${\cal D} < 0$.

\section{Numerical Estimates}

\subsection{Interference effect in MSSM}

Assuming the MSSM
as an example for various multi-Higgs-doublet models,
we present numerical estimates of the asymmetry ${\cal A}$.
The MSSM include three neutral Higgs bosons, 
two of which are CP-even and the other is CP-odd.
The Higgs sector can be parameterized by two parameters,
the mass of the CP-odd Higgs boson $m_A$ and
the ratio of the vacuum expectation values $\tan\beta$.
If $m_A$ is sufficiently large, the mass of the heavier
CP-even Higgs boson $m_H$ becomes approximately
degenerated into $m_A$.
For definiteness, we take $m_A=400\GEV$ and
$\tan\beta$ = 3, 7, 15 and 30.
The masses, the decay widths and the decay branching ratios of the 
Higgs bosons from which we derive the coupling constants
among Higgs and fermions
are computed by the program HDECAY~\cite{HDECAY},
which are listed in Table~\ref{masses}. For the input parameters
in the program, 
the sfermion mass scale is set for $m_{\mbox{\tiny SUSY}}$ = 1 TeV, 
the SU(2) gaugino mass parameter $M_2$ for 500 GeV and 
the higgsino mixing mass parameter $\mu$ for $-500$ GeV, 
which result in heavy supersymmetric particles.   
No new particles other than Higgs bosons are produced 
by the decay of the $H$ and $A$ bosons for the parameters. 

\begin{figure}[h]
\begin{center}
\epsfxsize=12cm
\epsffile{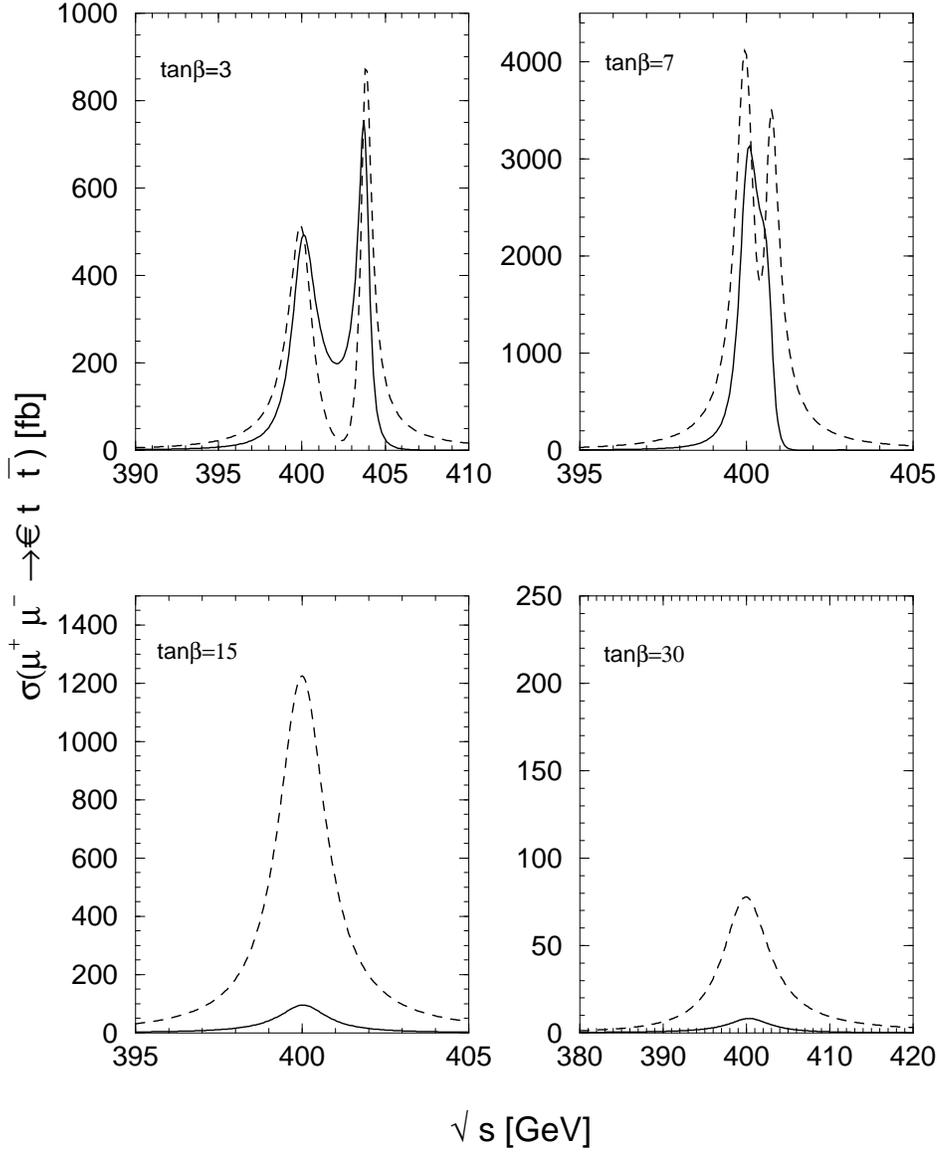}
\caption{The cross sections of $\mu^+\mu^- \rightarrow t\overline{t}$
with Higgs resonances for $\tan\beta=3,~7,~10$ and $30$.
The solid curves show the cross sections 
for $\mu^+_L \mu^-_L \rightarrow t_L \overline{t}_L$ or
$\mu^+_R \mu^-_R \rightarrow t_R \overline{t}_R$,
the dashed curves for
$\mu^+_L \mu^-_L \rightarrow t_R \overline{t}_R$ or
$\mu^+_R \mu^-_R \rightarrow t_L \overline{t}_L$.}\label{higgs}
\end{center}
\end{figure}

In Fig.~\ref{higgs} we
show the center-of-mass-energy dependence of 
the cross sections for 
the helicity combinations satisfying $\lambda_i=\lambda_f=0$.
The cross sections for $\Lambda=\overline{\Lambda}=\lambda=\overline{\lambda}$,
$\sigma^{LLLL}$ and $\sigma^{RRRR}$,
are different from those for 
$\Lambda=\overline{\Lambda} \ne \lambda=\overline{\lambda}$, $\sigma^{LLRR}$
and $\sigma^{RRLL}$.
These differences come from the interference effect between 
the Higgs-exchange amplitudes.
 The peak cross section is maximized at $\tan\beta \sim$  7, as 
$\sigma \simeq$ 4000 fb. 
The cross section mediated by $\gamma$- and $Z$-exchange diagrams are
smaller than 0.01 fb for $\lambda_i = \lambda_f = 0$, 
and generally negligible in these energy range.
A Large interference effect can be seen when the mass difference 
between $H$ and $A$ bosons is smaller than their widths
and two Higgs-exchange amplitudes have comparable magnitudes,
as is shown for $\tan\beta=15$ and $30$ in Fig.~\ref{higgs}.

The asymmetry ${\cal A}$ for the cross sections
is given in Table~\ref{asym-value} for 
several values of $\sqrt{s}$ at and around the resonances.
The asymmetry reaches to ${\cal O}(1)$ 
in certain ranges of $\sqrt{s}$, showing the strong effect of the 
interference between $H$ and $A$.
Since ${\cal A}$ is negative for 
$\sqrt{s} < m_{H,A}$, we can deduce
\begin{eqnarray}
a_{\mu} \cdot a_t \cdot b_{\mu} \cdot b_t < 0
\end{eqnarray}
which is consistent with eq.~(\ref{MSSMcup}).

\subsection{Beam polarization and helicity observation}

\begin{figure}[h]
\begin{center}
\epsfxsize=12cm
\epsffile{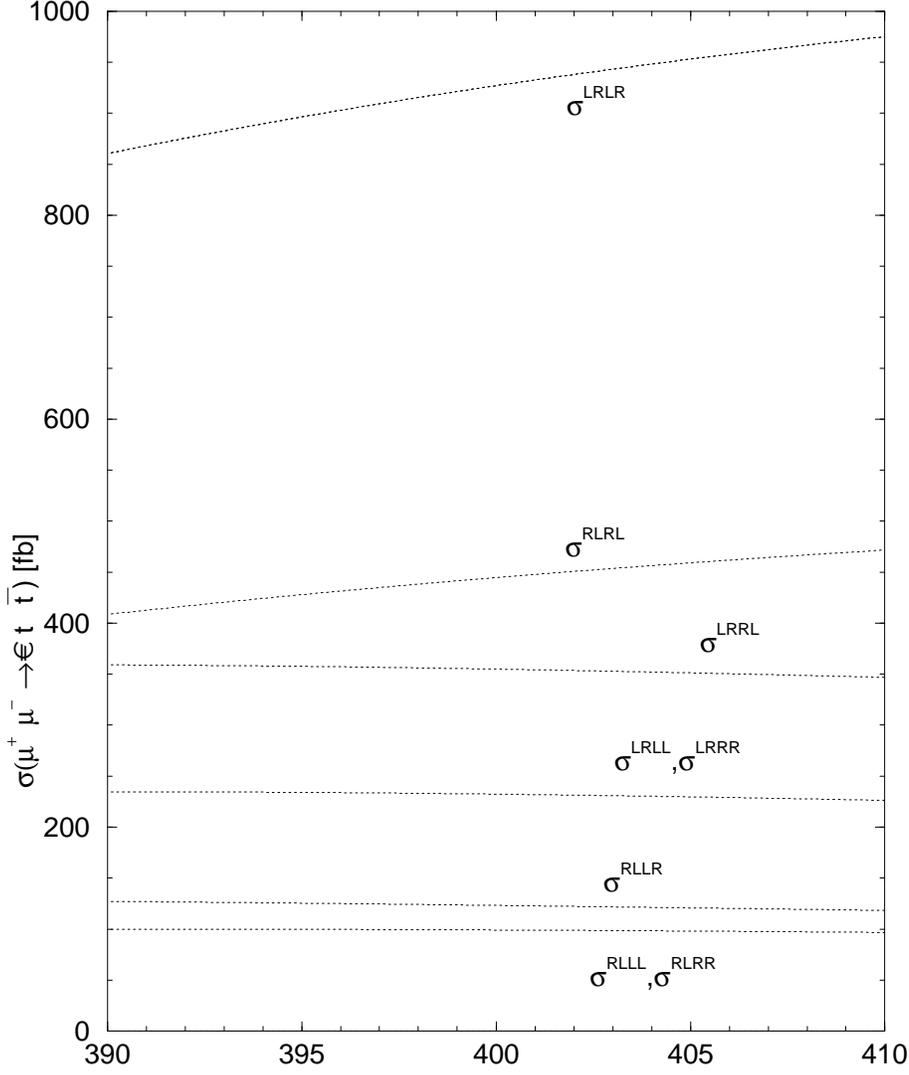}
\caption{The cross sections of $\mu^+\mu^- \rightarrow \gamma$/$Z \rightarrow
t\overline{t}$. The cross sections, $\sigma^{LLLL}$, $\sigma^{LLRR}$,
$\sigma^{RRRR}$, $\sigma^{RRLL}$, $\sigma^{LLLR}$, $\sigma^{LLRL}$,
$\sigma^{RRLR}$ and $\sigma^{RRRL}$ are smaller than 0.01 [fb].
} \label{gZ}
\end{center}
\end{figure}

The above arguments did not take account of the degree of 
polarization of the initial muon beams and 
the helicity observation of the final top pairs.
As was mentioned in section $2$,
the absolute values of the amplitudes of $\gamma$- and $Z$-exchange 
diagrams with helicity 
combinations other than $\lambda_i=\lambda_f=0$
are generally larger than
those with $\lambda_i=\lambda_f=0$ (Fig.~\ref{gZ}).
Therefore, the background processes 
$\mu^+ \mu^- \rightarrow \gamma/Z \rightarrow t \overline{t}$
should be considered as long as the polarization of muon beams and
the efficiency of helicity observation of top pairs are not
perfect.
The cross sections are contaminated with these backgrounds.

There are some statistical methods to measure the top-quark
helicity~\cite{AKSW,HMW,lepton}.
As an illustration, we follow Ref.~\cite{AKSW}.
The bottom quark decaying from a top quark has the angular
distribution proportional to $0.5 - 0.2 \lambda \cos\theta$,
where $\theta$ is the emission angle of the bottom
quark in the rest frame of the decaying top quark with respect
to the direction of the top momentum in the $t\overline{t}$
c.m.\ frame.
The anti-bottom quark has the distributon proportional to
$0.5 + 0.2 \overline{\lambda} \cos\overline{\theta}$,
where $\overline{\theta}$ is the emission angle of the anti-bottom
quark
\footnote{In order to derive the emission angles, $\theta$ and
$\overline{\theta}$, one needs to identify the charge and the momentum
of the top and the anti-top quarks.
The tasks can be done by observing their decay products.
For example, when one $W$ decays leptonically and the other
decays hadronically, we can generally reconstruct the top
and the anti-top decay kinematics.  Semileptonic decays of
the $B$ mesons can be used when both $W$'s decay hadronically.
The efficiency of the analysis of such $t\overline{t}$ reconstructions
is discussed in Ref.~\cite{snowmass}.}.
Therefore,
an event with the top quark decaying forward and with the anti-top
quark decaying backward is most likely to be a $t_L \overline{t}_L$
event.
Applying the emission angle cuts
$F$ (forward, $0 < \cos\theta, \cos\overline{\theta} < +1$)
and $B$ (backward, $-1 < \cos\theta, \cos\overline{\theta} < 0$)
to the bottom quark diretion,
the counted numbers of polarized top and anti-top decay events
are reduced by the factor $R$ and $\overline{R}$,
respectively.
\begin{equation}
\begin{array}{rclrcl}
R(F,\lambda) & = & \frac{\DS 1-2\lambda r}{\DS 2}, &
R(B,\lambda) & = & \frac{\DS 1+2\lambda r}{\DS 2}, \\
\overline{R}(F,\overline{\lambda}) & = &
\frac{\DS 1+2\overline{\lambda} r}{\DS 2}, &
\overline{R}(B,\overline{\lambda}) & = &
\frac{\DS 1-2\overline{\lambda} r}{\DS 2},
\end{array}
\end{equation}
where $r$ is $r_B-r_F$ in Ref.~\cite{AKSW}, and its value is
estimated to be 0.2 in this method.
If one observe the direction of the charged lepton from the
$t \rightarrow bW \rightarrow bl\nu$ decay~\cite{lepton},
instead of the bottom quark direction, $r$ reaches to be 0.5
in the Born approximation. 

\begin{figure}[h]
\begin{center}
\epsfxsize=12cm
\epsffile{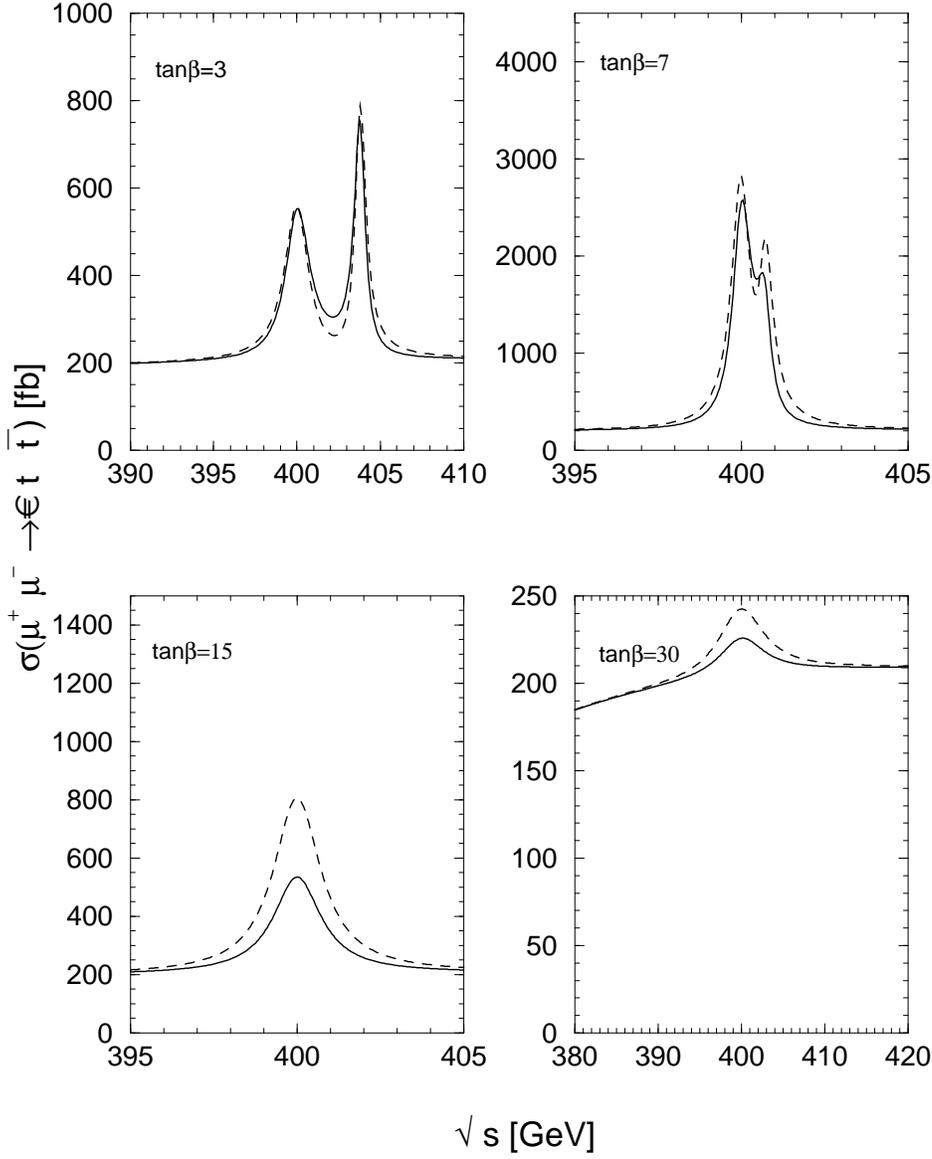}
\caption{The effective cross sections of $\mu^+\mu^-
\rightarrow t\overline{t}$
for $\tan\beta=3,~7,~10$ and $30$.
The solid curves are for $\sigma(P,P,B,F)$  or
$\sigma(-P,-P,F,B)$, while the dashed curves for
$\sigma(P,P,F,B)$ or $\sigma(-P,-P,B,F)$, with
$P=+0.6$ and $r=0.2$.  } \label{eff}
\end{center}
\end{figure}

We define the effective cross sections as follows:
\begin{eqnarray}
\sigma
(P_{\mu^-}, P_{\mu^+}, c, \overline{c})
= \sum\limits_{\Lambda \overline{\Lambda} \lambda \overline{\lambda}}
\sigma^{\Lambda \overline{\Lambda} \lambda \overline{\lambda}}
\frac{(1+2\Lambda P_{\mu^-})}{2} \frac{(1+2\overline{\Lambda} P_{\mu^+})}{2}
R(c,\lambda) \overline{R}(\overline{c},\overline{\lambda}),
\end{eqnarray}
where $P_{\mu^-}$ and $P_{\mu^+}$
denote the degree of polarizations of $\mu^-$ and $\mu^+$, and
$c$ and $\overline{c}$ are $F$ or $B$, respectively.
The effective cross sections are shown in Fig.~\ref{eff},
assuming $|P_{\mu^-}|=|P_{\mu^+}|=0.6$ as an example.
 To extract the Higgs signal efficiently, we should select as
$P_{\mu^-}=P_{\mu^+}=P$ and,
$c=F$ and $\overline{c}=B$ or $c=B$ and $\overline{c}=F$.
 In these selections, the background cross sections of $\gamma$
and $Z$ are independent on the signs of $P$'s and the choice of
$c$.
There is another interference contribution between the Higgs-exchange
and the $\gamma$- or $Z$-exchange amplitude
for $\lambda_i = \lambda_f = 0$,
and this interference has helicity dependence, however, it is
negligibly small.
With the polarizations and the emission angle cuts, we can define another
asymmetry of the cross sections between the different cuts,
with $P$ unchanged.
\begin{eqnarray}
{\cal A}^{'} &=& \frac{ \sigma(P,P,B,F) - \sigma(P,P,F,B) }
                      { \sigma(P,P,B,F) + \sigma(P,P,F,B) }
\nonumber \\
        &=& \frac{\sigma^{diff}}{\sigma^{tot}}
\end{eqnarray}
\begin{eqnarray}
\sigma^{diff} &=& \frac{\epsilon}{4} \Biggl\{
     (1+P)^2 \bigl( \sigma^{RRRR} - \sigma^{RRLL} \bigr)
   +(1-P)^2 \bigl( \sigma^{LLRR} - \sigma^{LLLL} \bigr)
\Biggr\}
\nonumber \\
\sigma^{tot} &=& \frac{1}{8} \Biggl\{
     (1+P)^2 \Bigl[ (1+\epsilon^2) \bigl( \sigma^{RRRR} + \sigma^{RRLL}
                  \bigr)
                   + (1-\epsilon^2) \bigl( \sigma^{RRRL} + \sigma^{RRLR}
\bigr)
             \Bigr]
\nonumber \\
  &+& ~~~~ (1-P^2) \Bigl[ (1+\epsilon^2) \bigl( \sigma^{LRRR} +
\sigma^{RLLL}
                                         + \sigma^{RLRR} + \sigma^{LRLL}
\bigr)
\nonumber \\
  && ~~~~~~~~~~~~~  + (1-\epsilon^2) \bigl( \sigma^{LRRL} + \sigma^{RLLR}
                                         + \sigma^{RLRL} + \sigma^{LRLR}
\bigr)
             \Bigr]
\nonumber \\
  &+& ~~~~ (1-P)^2 \Bigl[ (1+\epsilon^2) \bigl( \sigma^{LLRR} +
\sigma^{LLLL}
                   \bigr)
                   + (1-\epsilon^2) \bigl( \sigma^{LLRL} + \sigma^{LLLR}
\bigr)
             \Bigr] \Biggr\}
\end{eqnarray}
As is demonstrated in Table~\ref{asym-value},
the ratio of the asymmetries ${\cal A}/{\cal A}^{'}$ is at least three
and sometimes reaches to several ten, which means the strong suppression due
to the
$\gamma$ and $Z$ contributions.

Since the $\gamma$- and $Z$-exchange cross sections are well-known,
we can define the other asymmetry $\cal{A}^{''}$ with extracting the
$\gamma$ and $Z$ contributions:
\begin{eqnarray}
{\cal A}^{''} &=& \frac{ \sigma(P,P,B,F) - \sigma(P,P,F,B)
-\sigma^{diff}_{\gamma,Z}}
                       { \sigma(P,P,B,F) + \sigma(P,P,F,B)
-\sigma^{tot}_{\gamma,Z}},
\end{eqnarray}
where $\sigma^{diff}_{\gamma,Z}$ and $\sigma^{tot}_{\gamma,Z}$
are the $\gamma$ and $Z$ contributions in $\sigma^{diff}$ and
$\sigma^{tot}$,
respectively.
Here $\sigma^{diff}_{\gamma,Z}$ is vanishing.
Since we have extracted the background effects of
$\gamma$- and $Z$-exchange from ${\cal A}^{'}$,
the ${\cal A}^{''}$ is proportional to ${\cal A}$:
\begin{eqnarray}
{\cal A}^{''} = \frac{4}{\Bigl( \frac{1}{r}+r \Bigr)
                         \Bigl( \frac{1}{P}+P \Bigr)} ~ {\cal A}.
\end{eqnarray}
When ${\cal A}^{''} > 0.1 ~ {\cal A}$ is required,
$r > 0.3$ for $P=0.1$~\cite{precise-measure},
and $P > 0.13$ for $r=0.2$.
The values for $P=0.6$ and $r=0.2$ are shown in
Table~\ref{asym-value}.

Neglecting the systematic errors and assuming 100{\%}
efficiency, we found that required luminosity to 
establish the non-zero asymmetry ${\cal A}^{''}$ 
is least at moderate value of $\tan\beta$.  
Only $\simgt 20~ {\rm pb}^{-1}$ of the integreted 
luminosity enables us to perceive a
non-vanishing asymmetry within 1-$\sigma$
statistical error for $\tan\beta = 15$, and
$30~ {\rm pb}^{-1}$ for $\tan\beta = 7$.
For $\tan\beta = 3$ and $30$, $0.25~ {\rm fb}^{-1}$ and
$1.5~ {\rm fb}^{-1}$ are required, respectively.
%

\section{Conclusions}
We have discussed the interference effect on the cross section
of the process $\mu^+\mu^- \rightarrow t\overline{t}$
with $H$ and $A$ resonances almost degenerated.
The interference between $H$ and $A$ bosons arises if muon beams 
are longitudinally polarized and if we observe the helicities 
of the top quarks.
The interference effect can be measured by observing the difference
between the cross sections with appropriate helicity combinations.
It has been shown that the existence of the difference can be
reliable evidence of the existence of both $H$ and $A$ bosons,
while the absence indicates the existence of only $H$ bosons
or $A$ bosons around the resonances.
It is especially important that the existence of the $H$ and $A$ 
bosons can be established even if their masses are degenerated, 
by observing the interference effect in the overlapped resonances.  
We have estimated an asymmetry between the cross sections
adopting the MSSM as an example.
Even after taking into account 
the background $\gamma$- and $Z$-exchange contributions
because of incompleteness of the polarizations of muon beams 
and the measurement of top quark helicities,
the absolute values of the asymmetry can be detectable
in certain ranges of $\sqrt{s}$.
If we can accumulate the luminosities which enable us to perceive
the asymmetry, it is possible to learn not only the existence of 
$H$ and $A$ bosons but also the sign of the product of coupling constants
of Higgs bosons to fermions from the sign of the asymmetry.

\vspace{7mm}

{\bf Acknowledgments}
~The authors thank K.~Hagiwara and N.~Oshimo for valuable discussions
and reading manuscript. They also would like to thank
K.~Fujii, K.~Ikematsu and T.~Ohgaki for useful comments.
This work is supported in part by the Grant-in-Aid for
Scientific Research (No.~11640262) and the Grant-in-Aid
for Scientific Research on Priority Areas (No.~11127205)
from the Ministry of Education, Science and Culture, Japan.

\newpage
%
%
%
\newcommand{\Journal}[4]{{\sl #1} {\bf #2} {(#3)} {#4}}
\newcommand{\PL}{\sl Phys. Lett.}
\newcommand{\PR}{\sl Phys. Rev.}
\newcommand{\PRL}{\sl Phys. Rev. Lett.}
\newcommand{\NP}{\sl Nucl. Phys.}
\newcommand{\ZP}{\sl Z. Phys.}
\newcommand{\PTP}{\sl Prog. Theor. Phys.}
\newcommand{\NC}{\sl Nuovo Cimento}
\newcommand{\MPL}{\sl Mod. Phys. Lett.}
\newcommand{\PRep}{\sl Phys. Rep.}
\newcommand{\EPJ}{\sl Eur. Phys. J.}
\newcommand{\CPCR}{\sl Comm. Phys. Commun. Res.}

%
%
%
\newpage
\section*{Tables}

\begin{table}[h]
\caption{The helicity dependence of the Higgs amplitudes of 
$\mu^+ \mu^- \rightarrow H$/$A \rightarrow t \bar{t}$.
We denote ${\cal M}_{H/A}^{RRRR}$ as ${\cal M}_{H/A}$ for 
simplicity.
\label{helamp}}
\vspace{0.4cm}
\begin{center}
\begin{tabular}{|c|c|c|}
\hline
&&\\
&
{\large $t_L \bar{t}_L$}&
{\large $t_R \bar{t}_R$}
\\ 
\hline
&&\\
{\large $\mu^-_L \mu^+_L$}&
\begin{minipage}{1.0in}
\begin{center}
$~~{\cal M}_H $\\
$~~{\cal M}_A$
\end{center}
\end{minipage}
&
\begin{minipage}{1.0in}
\begin{center}
$-{\cal M}_H $\\
$~~{\cal M}_A$
\end{center}
\end{minipage}
\\ 
&&\\
\hline
&&\\
{\large $\mu^-_R \mu^+_R$}&
\begin{minipage}{1.0in}
\begin{center}
$-{\cal M}_H $\\
$~~{\cal M}_A$
\end{center}
\end{minipage}
&
\begin{minipage}{1.0in}
\begin{center}
$~~{\cal M}_H $\\
$~~{\cal M}_A$ 
\end{center}
\end{minipage}
\\ 
&&\\
\hline
\end{tabular}
\end{center}
\end{table}

\begin{table}[h]
\begin{center}
\caption{The masses, the total decay widths, 
the $\mu^+ \mu^-$ decay branching ratios and 
the $t\overline{t}$ decay branching ratios of 
the $H$ and $A$ bosons in the MSSM 
adopted in our numerical simulations.}
\vspace{7mm}
\begin{tabular}{|c||c|c|c|c|}
\hline
 $\tan\beta$ & $m_H$ & $\Gamma_H$ & 
               $Br(H \rightarrow \mu^-\mu^+)$ & 
               $Br(H \rightarrow t\overline{t})$ \\
 & (GeV) & (GeV) & $10^{-4}$ & $10^{-2}$ \\
\hline 
3.0 & 403.78 & 0.79 & 0.33 & 74.2 \\
7.0 & 400.71 & 0.50 & 2.84 & 20.7 \\
15.0& 399.97 & 1.70 & 3.88 & 1.30 \\
30.0& 399.49 & 6.67 & 3.95 & 0.078\\
\hline \hline
 $\tan\beta$ & $m_A$ & $\Gamma_A$ & 
               $Br(A \rightarrow \mu^-\mu^+)$ & 
               $Br(A \rightarrow t\overline{t})$ \\
 & (GeV) & (GeV) & $10^{-4}$ & $10^{-2}$\\
\hline
3.0 & 400.00 & 1.75 & 0.15 & 94.6 \\
7.0 & 400.00 & 0.67 & 2.13 & 45.2 \\
15.0& 400.00 & 1.74 & 3.80 & 3.83 \\
30.0& 400.00 & 6.69 & 3.95 & 0.25 \\
\hline 
\end{tabular}
\label{masses}
\end{center}
\end{table}

\begin{table}[h]
\begin{center}
\caption{The cross sections and the asymmetries.
We denote $\sigma_{H,A}(P,P,c,\overline{c})$ with
$\gamma$ and $Z$ contributions extracted
to be $\sigma_{H,A}(c,\overline{c})$, for simplicity,
where $P$ and $r$ are assumed to be $+0.6$ and $0.2$,
respectively, and $c$ and $\overline{c}$ are $F$
(forward) or $B$ (backward) of the angle cut.
}\label{asym-value}
\begin{tabular}{|c||r| r | r | r || r || r | r | r |}
\hline
&&&&&&&& \\
$\tan\beta$ & $\sqrt{s}~~~$ & $\sigma^{LLLL}$ & $\sigma^{LLRR}$
& ${\cal A}~~~~$ & ${\cal A'}~~~~$
& $\sigma_{H,A}(B,F)$
& $\sigma_{H,A}(F,B)$
& ${\cal A}^{''}~~~~$\\
& (GeV) & (fb) & (fb) &  &  & (fb) & (fb) & \\
\hline\hline
3.0 & 400.0 & 484.9 & 510.4 & $-0.026$
    & $-0.006$ & 347.5 & 353.5 & $-0.009$ \\
    & 402.5 & 213.8 & 23.66 & 0.801
    & 0.079 & 107.4 & 61.47 & 0.272 \\
    & 405.0 & 24.56 & 161.5 & $-0.736$
    & $-0.061$ & 48.29 & 80.69 & $-0.251$ \\
    & 406.0 & 3.260 & 70.60 & $-0.912$
    & $-0.035$ & 17.49 & 33.31 & $-0.311$ \\
\hline
7.0 & 399.0 & 152.2 & 591.2 & $-0.591$
    & $-0.113$ & 209.1 & 315.4 & $-0.203$ \\
    & 400.0 & 2969.1& 4071.7& $-0.157$
    & $-0.049$ & 2347.1 & 2610.4 & $-0.053$ \\
    & 400.5 & 2358.1& 1905.0&  0.106
    & 0.032 & 1557.6 & 1447.3 & 0.037\\
    & 401.5 & 1.860 & 684.7 & $-0.995$
    & $-0.183$ & 159.0 & 321.1 & $-0.338$ \\
\hline
15.0& 398.0 & 16.19 & 184.6 & $-0.839$
    & $-0.074$ & 50.74 & 91.60 & $-0.287$ \\
    & 400.0 & 95.22 & 1226.1& $-0.856$
    & $-0.202$ & 329.8 & 599.9 & $-0.291$ \\
    & 401.0 & 42.70 & 527.7 & $-0.850$
    & $-0.143$ & 142.5 & 257.4 & $-0.287$ \\
\hline
30.0& 390.0 & 0.88 & 6.39 & $-0.758$
    & $-0.003$ & 1.93 & 3.37 & $-0.272$ \\
    & 400.0 & 8.20 & 77.76& $-0.809$
    & $-0.036$ & 21.88 & 38.49 & $-0.275$ \\
    & 405.0 & 2.43 & 25.80& $-0.828$
    & $-0.013$ & 7.08 & 12.44 & $-0.275$ \\
\hline
\end{tabular}
\end{center}
\end{table}

%
%
\newpage

\end{document}